\documentclass[
reprint,
onecolumn,
nofootinbib,
amsmath,amssymb, aps
]{revtex4-2}
\pdfoutput=1

\usepackage[utf8]{inputenc}
\usepackage{graphicx}
\usepackage{amsmath}
\usepackage{amssymb}

\usepackage[usenames,dvipsnames]{xcolor}

\begin{document}

\title{Hybrid model of condensate and particle Dark Matter: linear perturbations in the hydrodynamic limit}
\author{Nick P. Proukakis}
\email[E-mail: ]{nikolaos.proukakis@newcastle.ac.uk}
\author{Gerasimos Rigopoulos}
\email[E-mail: ]{gerasimos.rigopoulos@newcastle.ac.uk}
\author{Alex Soto}
\email[E-mail: ]{alex.soto@newcastle.ac.uk}
\affiliation{School of Mathematics, Statistics and Physics, \\ Newcastle University, Newcastle upon Tyne, NE1 7RU, UK}
\date{June 2024}

\begin{abstract}
\noindent We analyse perturbations of self-interacting, scalar field dark matter that contains modes both in a coherent condensate state and an incoherent particle-like state. Starting from the coupled equations for the condensate, the particles' phase space distribution and their mutual gravitational potential, first derived from first principles in earlier work by the authors, we derive a hydrodynamic limit of two coupled fluids and study their linearized density perturbations in an expanding universe, also including particle pressure under an assumption for an equation of state consistent with the dynamical equations. We find that away from the condensate-only or particle-only limits, and for certain ranges of the parameters, such self-interacting mixtures can significantly enhance the density power spectrum above the standard linear $\Lambda$CDM value at localised wavenumbers, even pushing structure formation into the non-linear regime earlier than expected in $\Lambda$CDM for these scales. We also note that such mixtures can lead to degeneracies between models with different boson masses and self-coupling strengths, in particular between self-coupled models and non-coupled Fuzzy Dark Matter made up of heavier bosons. These findings open up the possibility of a richer phenomenology in scalar field dark matter models and could further inform efforts to place observational limits on their parameters.        
     
\end{abstract}

\maketitle

\section{Introduction}
For many decades now, Cold Dark Matter (CDM) has been the prevailing paradigm for explaining structure formation in the universe. Indeed, the assumption that the matter budget of the universe is dominated by a non-relativistic particle with small velocity dispersion (i.e.~``cold'') and negligible coupling to itself and the particles of the standard model has been able to explain a vast swathe of observations, spanning the whole history of the observable universe \cite{Frenk-White_DM}. Despite these successes however, more detailed investigations of the small scale structures predicted by numerical simulations of CDM have led to a surge in the scrutiny of an alternative dark matter model called Fuzzy Dark Matter (FDM) \cite{Hu:2000ke,Marsh:2010wq,Marsh:2013ywa,2014NatPh..10..496S,Schive:2014hza,Hlozek:2014lca,Schive:2015kza,Mocz:2017wlg,Veltmaat:2019hou,Du:2018qor,Mina:2020eik,Lague:2020htq,Hartman:2022cfh} - see \cite{Marsh:2015xka,Hui:2021, 2021A&ARv..29....7F} for detailed reviews and comprehensive bibliography - which promises to solve CDM's small scale problems such as the Cusp-Core problem \cite{Moore:1994yx}, the too-big-to-fail problem \cite{Boylan-Kolchin:2011qkt,Boylan-Kolchin:2011lmk} or the missing satellites problem \cite{Kauffmann:1993gv}. Although the extent to which these problems pose fundamental challenges to CDM is still unclear, their investigation, coupled with a lack of any compelling particle candidate for dark matter, has made FDM and its generalizations an attractive subject of study.

Unlike the corpuscular nature of CDM, with N-body simulations probing its phase space distribution, FDM is modelled as a continuous field  exhibiting wave interference effects. Although the resulting large scale structure is very similar to CDM, there are differences on smaller scales:  halos now exhibit a ``granular" appearance and FDM predicts the formation of compact objects from the balance of gravity and the ``quantum pressure" arising from that wavy dynamics. Such objects are the ground states of FDM and in a cosmological context reside in the centres of more extended halos. These features present a natural cutoff in the overdensity power spectrum of FDM precisely because the quantum pressure acts to prevent short scale overdensities from forming.       

Although sometimes associated to, or motivated by, speculative axion-like particles, FDM is in its basis a dark matter model composed of bosons, in contrast to CDM which is presumably composed of some unknown type of fermion. Bosons can exist in both condensed and non-condensed states, and FDM is usually associated with the former state - see however \cite{Liu:2022rss} for a quantitative assessment of the level of coherence, a related concept and a qualifier of condensation, inside an FDM halo and \cite{Brandenberger:2023idg} on how coherence of condensates of low mass particles formed in the early universe is prone to decay. Depending on how the dark matter bosonic field came into existence in the early universe, it may be that a particle description is also appropriate or at least offers a sufficiently accurate dynamical picture for some of the field's modes associated with higher velocities. In this work we utilize a formalism developed by us in \cite{Proukakis:2023nmm} to describe hybrid states, where the bosonic field can have both condensed and non-condensed components, and study linearized perturbations of such a model.\footnote{A similar dual particle-condensate description for homogeneous states in the early universe was recently put forward in \cite{Ai:2023qnr} using the 2PI formalism. It would be interesting to see how that approach relates to the one adopted in \cite{Proukakis:2023nmm}.} We work in the non-relativistic limit using a hydrodynamical truncation of the Boltzmann equation describing the non-condensed ``particles" to leading order in the self-coupling, ignoring collisional terms. 

The outline of the paper is as follows: In the next section we describe our basic equations and detail our assumptions and approximations, leading to a hydrodynamical, two-fluid description. The underlying physical picture is of the co-existence of two components, one involving states of relatively low velocities and being described by a classical scalar field $\Phi_0(t,\mathbf{r})$, and one with higher velocity states described by a phase space distribution $f(t,\mathbf{r},\mathbf{p})$. Linearized density perturbation equations for the two resulting fluids are then given in section \ref{sec3} and their solutions for the linear density power spectra are shown in section \ref{sec4}. The perturbation dynamics of the two components are essentially differentiated by the existence or not of a quantum pressure term $\propto k^4$. We observe that such coupled mixtures of condensed and non-condensed fluids, exhibit oscillatory features beyond the scale where FDM models present a quantum-pressure-induced cutoff and can exhibit an increase of power at certain localized wavenumbers, possibly pushing such scales into the non-linear regime at higher redshift compared to CDM. Of course, this invalidates perturbation theory and calls for fully non-linear simulations before any safe conclusions can be drawn on the effects of such models on structure formation. We also suggest that such mixtures may show degeneracies between models with different boson masses and self-coupling strengths, in particular between self-coupled models and non-self-interacting FDM ($g=0$) made up of \emph{heavier} bosons. Such degeneracies should be kept in mind when placing observational limits on models whose only parameter is the boson mass.

\section{Hybrid Model and Hydrodynamic equations}\label{sec2}

First we discuss our hybrid condensate-particle model and derive the corresponding hydrodynamic equations.

\subsection{Hybrid Condensate-Particle Model}

A generalised model encompassing both CDM and FDM as limiting cases was previously derived by us in~\cite{Proukakis:2023nmm}, based on the 
co-existence of two components for a single species of bosonic particle of mass $m$ under the effect of gravity. These two components are characterised by states of lower and higher typical velocities respectively. In flat spacetime the equations describing this system are
\begin{gather}
i \hbar  \frac{\partial \Phi_0}{\partial t}  = \left[ - \frac{\hbar^2}{2 m} \nabla^2 +
\left( m V + g (n_c + 2 \tilde{n}) \right) - i \hbar R \right]
\Phi_0 \label{initial1}\\
\frac{\partial f}{\partial t} + \frac{\mathbf{p}}{m} \cdot\nabla f - \nabla \left( m V + 2 g (n_c +
\tilde{n}) \right) \nabla_{\mathbf{p}}f = \frac{1}{2}
(I_a + I_b) \label{initial2}\\
\nabla^2 V = 4 \pi G m (n_c + \tilde{n}) \label{initial3}
\end{gather}
where equation \eqref{initial1} describes the part of the bosons that are in a condensate state (lower velocity modes), \eqref{initial2} is a collisional Boltzmann equation describing the incoherent particles, i.e. the non-condensed part (higher velocity modes), and \eqref{initial3} determines the gravitational potential generated by both of these components, whose number densities are defined as $n_c=|\Phi_0|^2$ for the condensate and $\tilde{n}$ for the incoherent particles; the latter's particle density is related with the distribution function $f$ which appears in \eqref{initial2} via
\begin{equation}\label{eq:num_density}
\tilde{n}= \int \frac{d^3 p}{(2 \pi \hbar)^3} f \;.
\end{equation}
The bosons are further allowed to self interact via a  self-interaction coupling $g$. The right-hand-side (rhs) of the Boltzmann equation involves the collision integrals
\begin{eqnarray}
I_a &=& 4 g^2 n_c \int \frac{d^3 p_1 d^3 p_2 d^3 p_3}{(2 \pi)^2 \hbar^4} \delta
(\varepsilon_{c} + \varepsilon_{\mathbf{p}_1} -
\varepsilon_{\mathbf{p}_2} - \varepsilon_{\mathbf{p}_3}) \delta
(\mathbf{p}_2 - \mathbf{p}_1 - \mathbf{q} + \mathbf{p}_3) \nonumber\\
& & \times(\delta
(\mathbf{p}_1 - \mathbf{p}) - \delta (\mathbf{p}_2 - \mathbf{p}) -
\delta (\mathbf{p}_3 - \mathbf{p})) \nonumber\\
& & \times((1 + f_1) f_2 f_3 - f_1 (1 + f_2) (1 + f_3))\\
I_b &=& 4 g^2 \int \frac{d^3 p_2 d^3 p_3 d^3 p_4}{(2 \pi)^5 \hbar^7} \delta
(\varepsilon_{\mathbf{p}_3} + \varepsilon_{\mathbf{p}_4} -
\varepsilon_{\mathbf{p}_2} - \varepsilon_{\mathbf{p}}) \delta
(\mathbf{p} + \mathbf{p}_2 - \mathbf{p}_3 - \mathbf{p}_4) \nonumber\\
& & \times [f_3 f_4 (f
+ 1) (f_2 + 1) - f f_2 (f_3 + 1) (f_4 + 1)]
\end{eqnarray}
where the subscripts in the distribution function are a shorthand $f_i=f(\mathbf{r},\mathbf{p}_i,t)$; here $\varepsilon_{c}$ corresponds to the energy of a condensate element with momentum $\mathbf{q}$, and the energies of the non-condensed particles, $\varepsilon_{\mathbf{p}}$, are respectively defined by
\begin{eqnarray}
\varepsilon_{c}=\frac{q^2}{2m}+ m V + g (n_c + 2 \tilde{n}) \;,\\
\varepsilon_{\mathbf{p}}=\frac{p^2}{2m}+ m V + 2g (n_c + \tilde{n}) \;.
\end{eqnarray}
Meanwhile, the condensate equation involves a damping term expressed in terms of 
\begin{equation}
R = \frac{1}{4 n_c} \int \frac{d^3 p}{(2 \pi \hbar)^3} I_a \label{eqr}    \;.
\end{equation}

Note that the effect of self-interaction is slightly different for the two components: there is a difference between the factors of $2$ in those terms on the rhs of \eqref{eqhyd2b} and \eqref{eqhyd4b} which are proportional to $g$. This originates in the different mean-field interaction potentials exhibited by condensed and non-condensed particles, and is well-known in condensed matter systems~\cite{proukakis2008finite,griffin_nikuni_zaremba_2009}. Notice further that in the limit $g\to 0$ and considering only the condensed component with $f=0$, we recover the limit of Fuzzy Dark Matter (FDM), while if $\Phi_0 = 0$ and we only have non-condensed particles, we recover the limit of the equations that are typically used for Cold Dark Matter (CDM). In this hybrid model, we can see that at order $g^2$ it is possible to have interchange of particles and condensate reflected in the existence of $-i \hbar R$ in \eqref{initial1} and $I_a$ in \eqref{initial2}. In this way, the self-interaction is a significant part of the phenomenology of the model. Even if we ignore these particle interchanges and only consider phenomena up to order $g$, the coupling between condensed and non-condensed components is still non-trivial.

We next map our equations \eqref{initial1}-\eqref{initial3} into a system of hydrodynamic equations. 

\subsection{Hydrodynamic Equations} 

The hydrodynamic equations for the condensate are routinely obtained via the Madelung transformation $\Phi_0 = \sqrt{n_c} e^{i \theta}$ \cite{Suarez:2011yf,Chavanis_2011a}: Splitting the resulting equation in real and imaginary parts and defining the condensate velocity $\mathbf{v}_c = \frac{\hbar \nabla \theta}{m}$, and with $\rho_c=m n_c$ and $\tilde{\rho}=m \tilde{n}$ being the mass densities for the condensate and particles respectively, we get for the condensate 
\begin{eqnarray}
\frac{\partial \rho_c}{\partial t} + \nabla \cdot (\rho_c \mathbf{v}_c) &=& - 2 \rho_c R \label{eqhyd1} \;, \\
\left( \frac{\partial}{\partial t} + \mathbf{v}_c \cdot \nabla \right)
\mathbf{v}_c &=& - \nabla \left( - \frac{\hbar^2}{2 m^2} \frac{\nabla^2
\sqrt{\rho_c}}{\sqrt{\rho_c}} + V + \frac{g}{m^2} (\rho_c + 2 \tilde{\rho})
\right) \label{eqhyd2} \;.
\end{eqnarray}
For the particles on the other hand, we can compute from equation \eqref{initial2} the first three moments, multiplying it by $1$, $(\mathbf{p} - m \mathbf{v}_c)$ and $(\mathbf{p} - m \mathbf{v}_c)^2$ respectively, and integrating with respect to $\mathbf{p}$. It can be verified that
\begin{eqnarray}
\int \frac{d^3 p}{(2 \pi \hbar)^3} I_b = 0, \quad \int \frac{d^3 p}{(2 \pi \hbar)^3} (\mathbf{p} - m \mathbf{v}_c) I_a = 0, \quad \int \frac{d^3 p}{(2 \pi \hbar)^3} (\mathbf{p} - m \mathbf{v}_c) I_b = 0\\
\int \frac{d^3 p}{(2 \pi \hbar)^3} (\varepsilon_{\mathbf{p}} - \varepsilon_c) I_a = 0, \qquad \int \frac{d^3 p}{(2 \pi \hbar)^3} \varepsilon_{\mathbf{p}} I_b = 0
\end{eqnarray}
and with these identities and Eq.~\eqref{eqr}, we obtain 
\begin{eqnarray}
\frac{\partial \tilde{\rho}}{\partial t} + \nabla \cdot (\tilde{\rho}
\widetilde{\mathbf{v}}) &=& 2 \rho_c R \label{eqhyd3}\\
\left( \frac{\partial}{\partial t} + \widetilde{\mathbf{v}} \cdot \nabla
\right) \widetilde{\mathbf{v}} &=& - \nabla \left( V + \frac{2 g}{m^2} (\rho_c
+ \tilde{\rho}) \right) \nonumber\\
& & - \frac{1}{\tilde{\rho}} \nabla_i P_{i j} - 2
\frac{\rho_c}{\tilde{\rho}} R (\widetilde{\mathbf{v}} - \mathbf{v}_c) \label{eqhyd4}\\
\frac{\partial E}{\partial t} + \nabla \cdot (E \widetilde{\mathbf{v}}) &=&
- \nabla \cdot \mathbf{Q} - D_{i j} P_{i j} \nonumber\\
& & + 2 \rho_c R \left( \frac{1}{2}
(\widetilde{\mathbf{v}} - \mathbf{v}_c)^2 - \frac{\hbar^2 \nabla^2
\sqrt{\rho_c}}{2 m^2 \sqrt{\rho_c}} - \frac{g}{m^2} \rho_c \right) \label{eqhyd5}
\end{eqnarray}
with $D_{i j} = \frac{1}{2} (\nabla_i \tilde{v}_j + \nabla_j \tilde{v}_i)$ and where we have defined as $\widetilde{\mathbf{v}}$ the bulk velocity of the particles as
\begin{equation}
\widetilde{\mathbf{v}} = \int \frac{d^3 p}{(2 \pi \hbar)^3}
\frac{\mathbf{p}}{m} \frac{f}{\tilde{n}}
\end{equation}
and
\begin{eqnarray}
P_{i j} &=& m \int \frac{d^3 p}{(2 \pi \hbar)^3} \left( \frac{p_i}{m} - \tilde{v}_i
\right) \left( \frac{p_j}{m} - \tilde{v}_j \right) f\\
E &=& \frac{1}{2 m} \int \frac{d^3 p}{(2 \pi \hbar)^3} (\mathbf{p} - m
\widetilde{\mathbf{v}})^2 f \\
\mathbf{Q} &=& \frac{1}{2 m} \int
\frac{d^3 p}{(2 \pi \hbar)^3} (\mathbf{p} - m \widetilde{\mathbf{v}})^2 \left(
\frac{\mathbf{p}}{m} - \widetilde{\mathbf{v}} \right) f
\end{eqnarray}
It is important to remark that, as is usually the case, the Boltzmann equation leads to an infinite hierarchy of moment equations and here we have only displayed equations for the first two moments. Accordingly, the existence of terms such as $\mathbf{Q}$ and $E$ shows that the above equations are not closed, requiring the equations for the next, cubic moments. Those in turn, would involve quartic moments, etc, in a never-ending hierarchy of equations. To truncate it, further assumptions are required.

A common assumption is that of local thermal equilibrium where $f$ assumes the form of the Maxwell-Boltzmann distribution with a locally varying temperature. Here we will not assume thermal equilibrium in particular but will assume that $f$ is an even function of $\mathbf{p} -  m \widetilde{\mathbf{v}} \equiv m\mathbf{u}$ 
\begin{equation}
    f=f(m\mathbf{u}) = f(-m\mathbf{u})\,,
\end{equation}
meaning that particle velocities are distributed symmetrically in frames comoving with the local average bulk velocity. With this symmetry assumption , $\mathbf{Q}=0$ and, for $P_{i j}$ the only non-zero entries are the diagonal ones:
\begin{equation}\label{eq:pressure}
P_{i i} = m \int \frac{d^3 p}{(2 \pi \hbar)^3} \left( \frac{p_i}{m} - \tilde{v}_i \right)^2 f \;. 
\end{equation}

Also, we will assume that the three diagonal entries are equal: $P_{11} = P_{22} = P_{33} \equiv P$, i.e. isotropy in the $P_{i j}$ term. With that, adding the three components
\begin{equation}\label{eq:pressure}
3P=\tilde{\rho}\left\langle\mathbf{u}^2\right\rangle = \Tilde{\rho} \mathbf{u}_\star^2\;,
\end{equation}

where 
\begin{equation}
\left\langle\mathbf{u}^2\right\rangle \equiv \frac{\int \frac{d^3 p}{(2 \pi \hbar)^3} \left( \frac{\mathbf{p}}{m} - \mathbf{\tilde{v}} \right)^2 f}{\int \frac{d^3 p}{(2 \pi \hbar)^3} f}\;.
\end{equation}
Here $\mathbf{u}^2_\star$ is a measure of the characteristic velocity squared, as determined by the distribution function $f$, measured in a local frame comoving with a fluid element. At the same time, with the isotropy assumption we have 
\begin{equation}
P = \frac{2}{3} E \;.
\end{equation}
a common relation for the kinetically induced pressure of a gas. Hence, equation \eqref{eqhyd5} becomes
\begin{equation}
\frac{\partial P}{\partial t} + \nabla \cdot (P \widetilde{\mathbf{v}}) =
- \frac{2}{3} P \nabla \cdot \widetilde{\mathbf{v}} + \frac{4}{3} \rho_c R
\left( \frac{1}{2} (\widetilde{\mathbf{v}} - \mathbf{v}_c)^2 -
\frac{\hbar^2 \nabla^2 \sqrt{\rho_c}}{2 m^2 \sqrt{\rho_c}} - \frac{g}{m^2} \rho_c \right) \; .
\end{equation}
We observe that these assumptions \emph{close the system of hydrodynamic equations}.

We highlight here that in order to construct the quantum pressure term $- \frac{\hbar^2}{2 m^2} \frac{\nabla^2 \sqrt{\rho_c}}{\sqrt{\rho_c}}$ in \eqref{eqhyd2} we needed to divide by $\sqrt{\rho_c}$ in some point, meaning that we can never take $\rho_c$ as zero. Also, to arrange equation \eqref{eqhyd4} in a similar form to \eqref{eqhyd2}, we divided by $\tilde{\rho}$. In this way, we can never take $\tilde{\rho}$ as zero either in the above coupled system of equations for the two components. If we want to consider the hydrodynamical equations for only condensate or only particles we should start from Eqs. \eqref{initial1} - \eqref{initial3}, setting to zero the absent component there and then repeat the steps for the remaining component.

Next, we formulate the above equations in the setting of an expanding universe. 

\subsection{Formulation in an Expanding Universe}
\label{IIC}

At this point, in the interest of simplicity,
we will keep the next computations at order $g$,  meaning that henceforth we set $R=0$ and, correspondingly, $I_a=I_b=0$. Including cosmic expansion we have that $\mathbf{r}=a \mathbf{x}$, where $\mathbf{r}$ is the physical coordinate, $\mathbf{x}$ the comoving coordinate and $a(t)$ the scale factor. Also, we have that the velocity is described by $\mathbf{u} = H \mathbf{r} + \mathbf{v}$, where $\mathbf{u}$ is the physical velocity, $ \mathbf{v}$ the peculiar velocity and $H=\dot{a}/a$. Considering the expansion, the time derivatives for a function can be written as $\left.\frac{\partial}{\partial t} \right |_{\mathbf{r}} = \left.\frac{\partial}{\partial t} \right |_{\mathbf{x}} - H \mathbf{x} \cdot \nabla$, where the nabla symbol stands for derivatives on the comoving coordinate $\mathbf{x}$. With this, the hydrodynamical equations in an expanding universe up to order $g$ are
\begin{gather}
\frac{\partial \rho_c}{\partial t} + 3 H \rho_c + \frac{1}{a} \nabla \cdot
(\rho_c \mathbf{v}) = 0 \label{eqhyd1b}\\
\frac{\partial \mathbf{u}}{\partial t} + \frac{1}{a} \mathbf{v} \cdot
\nabla \mathbf{u} = - \nabla \left( - \frac{\hbar^2}{2 m^2 a^3}
\frac{\nabla^2 \sqrt{\rho_c}}{\sqrt{\rho_c}} + \frac{1}{a} V + \frac{g}{m^2 a}
(\rho_c + 2 \tilde{\rho}) \right) \label{eqhyd2b}\\
\frac{\partial \tilde{\rho}}{\partial t} + 3 H \tilde{\rho} + \frac{1}{a}
\nabla \cdot (\tilde{\rho} \widetilde{\mathbf{v}}) = 0 \label{eqhyd3b}\\
\frac{\partial \widetilde{\mathbf{u}}}{\partial t} + \frac{1}{a}
\widetilde{\mathbf{v}} \cdot \nabla \widetilde{\mathbf{u}} = - \nabla
\left( \frac{1}{a} V + \frac{2 g}{m^2 a} (\rho_c + \tilde{\rho}) \right) -
\frac{1}{a \tilde{\rho}} \nabla P \label{eqhyd4b}\\
\frac{\partial P}{\partial t} + 5 H P + \frac{1}{a} \nabla \cdot
(\widetilde{\mathbf{v}} P) = - \frac{2}{3 a} P \nabla \cdot
\widetilde{\mathbf{v}} \label{eqhyd5b}\\
\nabla^2 V = 4 \pi G a^2 (\rho_c + \tilde{\rho}) \label{eqhyd6b}
\end{gather}
The above closed set of equations corresponds to a possible starting point for e.g. a perturbative analysis. 

Another simplification is possible if we further assume that the pressure of particles can be effectively described by an equation of state of the general form
\begin{equation}\label{eq:rhopart}
P=\kappa \tilde{\rho}^{n}
\end{equation}
for some value of $n$. We take $\kappa$ to be a constant, implicitly assuming that the underlying phase space distribution $f$ is quasi-static, at least on the timescales of interest; note that, as thermodynamic equilibrium is not established, $f$ \emph{need not} be a thermal distribution. Equation \eqref{eqhyd5b} then gives
\begin{equation}
\kappa\tilde{\rho}^{n - 1} \left[ n \frac{\partial \tilde{\rho}}{\partial t} + 5 H
\tilde{\rho} + \frac{5}{3} \frac{1}{a} \tilde{\rho} \nabla \cdot
\widetilde{\mathbf{v}} + \frac{1}{a} n \widetilde{\mathbf{v}} \cdot
\nabla \tilde{\rho} \right] = 0
\end{equation}
which, after the use of \eqref{eqhyd3b} leads to
\begin{equation}
\kappa\tilde{\rho}^n \left( \frac{5}{3} - n \right) \left( 3 H + \frac{1}{a} \nabla
\cdot \widetilde{\mathbf{v}} \right) = 0
\end{equation}
Hence, the general assumption \eqref{eq:rhopart} leaves us with four possibilities:
\begin{enumerate}
\item $\kappa=0$ which implies that the velocity dispersion pressure of the particle component is zero.
\item $\tilde{\rho}=0$ if $n>0$, in which case there is no particle component.  
\item $n=\frac{5}{3}$: This value is consistent with the dimension of $P$ and $n$. We assume that there is an effective $\kappa$ which can be approximated as a constant in position and time for the scales relevant to our computations which in principle can be constrained by observation.  

\item $3 H + \frac{1}{a} \nabla
\cdot \widetilde{\mathbf{v}}=0$  Since $\tilde{\mathbf{v}}$ is a perturbation this relation is impossible to satisfy in an expanding universe as it requires $3H = 0$ and implies no velocity perturbations. 
\end{enumerate}
Discarding possibility 4, we will explore all other cases. Note that the $n=5/3$ exponent is commonly discussed in the context of standard equilibrium statistical mechanics for non-interacting, monoatomic gases where the coefficient $\kappa$ is also computed as a function of the temperature and the chemical potential, i.e.~the parameters that enter the thermal equilibrium form of the distribution $f$. It is interesting to note that the exponent $5/3$ is also implied on dimensional grounds from the powers of momentum appearing in the defining integrals \eqref{eq:num_density} and \eqref{eq:pressure}, beyond thermal equilibrium assumptions. 
We stress again that we do not assume thermal equilibrium and the occupation of different modes need not be, and in fact is not, given by the equilibrium Bose-Einstein distribution. We will therefore treat $\kappa$ as a free parameter in the ensuing analysis, chosen such that it is compatible with the observed power spectrum.\footnote{It is worth noting that the equation of state \eqref{eq:rhopart} has been considered in \cite{Ahn:2004xt}, also assuming the same symmetries in $f$ and $P_{i j}$. For an interesting argument that a non-interacting, bosonic gas in local thermal equilibrium can provide viable halo models see \cite{2023arXiv231015795A}.} The general relation \eqref{eq:pressure} along with \eqref{eq:rhopart} gives
\begin{equation}\label{eq:typ_vel}
    u_\star = \sqrt{3\kappa} \tilde{\rho}^{1/3}
\end{equation}
for the characteristic random velocity of particles in the non-condensed component. 

We observe that according to our equations, the main distinction between the condensate and the particles lies in that the condensate possesses a quantum pressure term (first term on the rhs of \eqref{eqhyd2b}) while the particles exhibit a different pressure, given here by \eqref{eq:rhopart} and appearing on the rhs of \eqref{eqhyd4b}, stemming from phase space velocity dispersion. These two terms scale differently with wavenumber $k$, the first being proportional to $k^4$ while the second to $k^2$, see below \eqref{eqtosol1} and \eqref{eqtosol2}. We do not further speculate on the origins of these two different components but comment on the self-consistency of such a separation of phases below. Note that in assessing the relative abundance of the coherent and incoherent states generic measures of field coherence, such as those used in \cite{Liu:2022rss}, should be employed.

\section{Linear Regime equations}\label{sec3}
We will now study the perturbations around background values for the equations \eqref{eqhyd1b}-\eqref{eqhyd4b} and \eqref{eqhyd6b} along with \eqref{eq:rhopart}. Thus, considering the perturbations at first order 
\begin{equation}
\rho_c = \bar{\rho}_c (1 + \delta_c), \hspace{1.0cm} \tilde{\rho} = \bar{\rho}_{nc} (1 + \delta_{nc}), \hspace{1.0cm} V = V_0 + \delta V, \hspace{1.0cm} P = P_0 + \delta P
\nonumber
\end{equation}
and the peculiar velocities around zero, assuming the background follows the standard cosmological expansion of CDM, using the order zero equations in the first order ones and rearranging them,  we have that the equations for the condensate and particles density contrasts are
\begin{eqnarray}
& & \ddot{\delta}_c + 2 H \dot{\delta}_c + \left( \frac{\hbar^2 k^4}{4 m^2 a^4} -
4 \pi G \bar{\rho} f + \frac{g \bar{\rho} f k^2}{m^2 a^2} \right) \delta_c \nonumber\\
& & \hspace{1.0cm} - (1 - f) \left( 4 \pi G \bar{\rho} - \frac{2 g \bar{\rho} k^2}{m^2 a^2} \right)
\delta_{nc} = 0 \label{pasa}\\
& & \ddot{\delta}_{nc} + 2 H \dot{\delta}_{nc} - \left( 4 \pi G
\bar{\rho} (1 - f) -  \left( \frac{2 g \bar{\rho} (1 - f)}{m^2} +
\frac{5 \kappa \bar{\rho}^{2 / 3} (1 - f)^{2 / 3}}{3} \right) \frac{k^2}{a^2} \right)
\delta_{nc} \nonumber\\
& & \hspace{1.0cm} - f \left( 4 \pi G \bar{\rho} - \frac{2 g \bar{\rho}
k^2}{m^2 a^2} \right) \delta_c = 0 \label{pasb}
\end{eqnarray}
where we defined the effective condensate fraction $f$ as\footnote{Not to be confused with the phase space density of the non-condensed component used in section II.}
\begin{equation}\label{eq:cond-frac}
f = \frac{\bar{\rho}_c}{\bar{\rho}}
\end{equation}
and
\begin{equation}
\label{sumofrhobar}
\bar{\rho}=\bar{\rho}_c + \bar{\rho}_{nc} \,.
\end{equation}
We stress here that since we are not assuming any sort of thermal equilibrium, we consider the condensate fraction \eqref{eq:cond-frac} as a free parameter, {\em a priori} unrelated to some fixed value of the particle phase space density and/or determined by a critical temperature. Instead we consider two populations of the bosons with different velocity distributions, one of which is cold enough (i.e. with low typical velocities) for wave phenomena to be evident and the other comprising of faster moving particles. We provide some estimates of these typical velocities for the parameters we consider below.

Note that taking the limit $g \rightarrow 0$ and $\kappa\to 0$ we obtain the following reduced equations
\begin{eqnarray}
\ddot{\delta}_c + 2 H \dot{\delta}_c + \left( \frac{\hbar^2 k^4}{4 m^2 a^4} -
4 \pi G \bar{\rho} f \right) \delta_c - 4 \pi G \bar{\rho} (1 - f)
\delta_{nc} = 0 \;,\\
\ddot{\delta}_{nc} + 2 H \dot{\delta}_{nc} - \bar{\rho} (1 - f)
4 \pi G \delta_{nc} - 4 \pi G \bar{\rho} f \delta_c = 0 \;.
\end{eqnarray}
These equations are formally identical to the equations used in~\cite{Schwabe:2020eac}, which however considered a different physical scenario of two different particle species making up the dark matter, one of which is light enough to exhibit wave-like properties on the relevant scales and the other being sufficiently massive such that it can be described by N-body simulations. This is physically distinct to our setting where both coherent and incoherent particles stem from an internal partition of the same field on the basis of phase-space density, as previously discussed in~\cite{Proukakis:2023nmm,Liu:2022rss}. In our general equations, setting $\kappa\to 0$ means that we are considering the non-condensate overdensity as a CDM-like pressureless fluid. Thus, a hybrid model with $\kappa=0$ describes the coexistence of a condensate with a bath of CDM-like particles. A non-zero $\kappa$ value generalises this situation, modelling the pressure arising from the velocity dispersion of the self-interacting particles, and allowing new possible effects in the non-condensate sector.

The equations can be manipulated a little bit further by making use of the Friedmann equation
\begin{equation}
H^2 = \frac{8 \pi G}{3} \left(\bar{\rho} +\rho_\gamma\right) + \frac{\Lambda}{3}
\end{equation}
in the form
\begin{equation}
1 = \bar{\Omega} + \Omega_{\gamma} + \Omega_{\Lambda}
\end{equation}
where $\bar{\Omega} = \frac{8 \pi G (\bar{\rho}_c+\bar{\rho}_{nc})}{3 H^2}$ is the total matter density parameter, composed by the mixture of the condensed and the non-condensed particles defined by our equations, 
$\Omega_{\gamma} = \frac{8 \pi G \bar{\rho}_{\gamma}}{3 H^2}$ is the radiation density
parameter, and $\Omega_{\Lambda} = \frac{\Lambda}{3 H^2}$ the cosmological
constant density parameter. We have assumed the baryonic matter to be negligible compared with dark matter. With these definitions we replace the density $\bar{\rho} = \frac{3 H^2 \bar{\Omega}}{8 \pi
G}$ in \eqref{pasa} and \eqref{pasb} and change the time variable to be the scale factor, obtaining
\begin{eqnarray}
& & \frac{d^2 \delta_c}{d a^2} + \left( \frac{3}{a} + \frac{d \ln H}{d a} \right)
\frac{d \delta_c}{d a} - \frac{3 \bar{\Omega}}{2 a^2} \left( f - \frac{\hbar^2
k^4}{6 H^2 \bar{\Omega} m^2 a^4} - \frac{g f k^2}{4 \pi G m^2 a^2} \right)
\delta_c \nonumber\\
& & \hspace{1.0cm} - \frac{3 \bar{\Omega}}{2 a^2} (1 - f) \left( 1 - \frac{g k^2}{2 \pi
G m^2 a^2} \right) \delta_{nc} = 0 \;, \label{master1}\\
& & \frac{d^2 \delta_{nc}}{d a^2} + \left( \frac{3}{a} + \frac{d \ln H}{d
a} \right) \frac{d \delta_{nc}}{d a} \nonumber\\
& & \hspace{1.0cm} - \frac{1}{a^2} \left( \frac{3
\bar{\Omega}}{2} (1 - f) - \frac{1}{8 \pi G a^2} \left( \frac{6 \bar{\Omega} g
(1 - f)}{m^2} + \frac{5 \kappa (1 - f)^{2 / 3} \bar{\Omega}^{2 / 3} (8 \pi
G)^{1 / 3}}{(3 H^2)^{1 / 3}} \right) k^2 \right) \delta_{nc} \nonumber\\
& & \hspace{1.0cm} - \frac{3 \bar{\Omega}}{2 a^2} f \left( 1 - \frac{g k^2}{2 \pi G m^2 a^2} \right)
\delta_c = 0 \;. \label{master2}
\end{eqnarray}
Note that the wavenumber enters \eqref{master2} as $k^2$ only, i.e.~like a typical pressure term, while \eqref{master1} contains the $k^4$ term characteristic of the quantum pressure felt by the condensate but not by the incoherent particles. The two density contrasts $\delta_c$ and $\delta_{nc}$ are coupled via terms proportional to the condensate fraction $f$ and $(1-f)$. Equations \eqref{master1} and \eqref{master2} will form the basis of the analysis that follows.

\section{Solution of the density contrasts in a matter dominated universe}\label{sec4}
Consider first the case of a radiation dominated universe with a cosmological constant, where
$\bar{\Omega} \approx 0$. Our equations \eqref{master1} and \eqref{master2} read
\begin{eqnarray}
\frac{d^2 \delta_c}{d a^2} + \left( \frac{3}{a} + \frac{d \ln H}{d a} \right)
\frac{d \delta_c}{d a} + \frac{\hbar^2 k^4}{4 H^2 m^2 a^6} \delta_c &=& 0 \;,\\
\frac{d^2 \delta_{nc}}{d a^2} + \left( \frac{3}{a} + \frac{d \ln H}{d
a} \right) \frac{d \delta_{nc}}{d a} &=& 0 \;.
\end{eqnarray}
We observe that $\delta_c$ and $\delta_{nc}$ are decoupled and the
self-interaction and particle pressure don't play any role. The fact that both components are decoupled and both solutions in this stage are known will be important below when we consider the transfer functions used in the literature for both components.

Now, we move to the matter dominated case, where both components are coupled and the behaviour is more involved. Here, $\bar{\Omega} \approx 1$ and our equations \eqref{master1} and \eqref{master2} read
\begin{eqnarray}
& & \frac{d^2 \delta_c}{d a^2} + \frac{3}{2 a} \frac{d \delta_c}{d a} -
\frac{3}{2 a^2} \left( f - \frac{\hbar^2 k^4}{6 H_0^2 \bar{\Omega}_0 m^2 a} -
\frac{g}{m^2} \frac{f k^2}{4 \pi G a^2} \right) \delta_c \nonumber\\
& & \hspace{1.0cm} - \frac{3}{2 a^2} (1 - f) \left( 1 - \frac{g}{m^2} \frac{k^2}{2 \pi G a^2} \right)
\delta_{nc} = 0 \;, \label{eqtosol1}\\
& & \frac{d^2 \delta_{nc}}{d a^2} + \frac{3}{2 a} \frac{d
\delta_{nc}}{d a} - \frac{3}{2 a^2} \left( (1 - f) - \left(
\frac{g}{m^2} \frac{(1 - f)}{2 \pi G a^2} + \frac{5 \kappa}{6 a} \left(
\frac{(1 - f)^2}{3 \pi^2 G^2 H_0^2 \bar{\Omega}_0} \right)^{\frac{1}{3}}
\right) k^2 \right) \delta_{nc} \nonumber\\
& & \hspace{1.0cm} - \frac{3}{2 a^2} f \left( 1 -
\frac{g}{m^2} \frac{k^2}{2 \pi G a^2} \right) \delta_c = 0 \;, \label{eqtosol2}
\end{eqnarray}
where we have used that in this case
\begin{equation}
H^2 = \frac{H_0^2 \bar{\Omega}_0}{a^3}
\end{equation}
and the values $H_0$ and $\bar{\Omega}_0$ are the values measured today.

We will solve these equations numerically. To set the initial conditions we can use that since condensate and particles are decoupled before the time of radiation-matter equality, and they each obey known equations individually, we can use the same transfer functions for $\delta_c$ and $\delta_{nc}$ as have been used to study them separately in the literature. Thus, for the non-condensed part $\delta_{nc}$ we use the typical power spectrum for $\Lambda$CDM
\begin{equation}
P (k) = (T (k))^2 A_s k^{n_s}
\end{equation}
where $A_s k^{n_s}$ is the primordial power spectrum with the values for $A_s$ and $n_s$ obtained from the PLANCK data \cite{Planck:2018vyg}, and we will use the following fitted transfer function $T(k)$ \cite{Bardeen:1985tr} defined by
\begin{equation}
T (k) = \frac{\ln \left( 1 + 2.34 \frac{k}{k_{eq}} \right)}{2.34
\frac{k}{k_{eq}}} \left( 1 + 3.89 \frac{k}{k_{eq}} + \left( 16.1
\frac{k}{k_{eq}} \right)^2 + \left( 5.46 \frac{k}{k_{eq}}
\right)^3 + \left( 6.71 \frac{k}{k_{eq}} \right)^4 \right)^{- \frac{1}{4}}
\end{equation}
where $k_{eq}$ is the wavenumber associated to the mode crossing the horizon in the radiation-matter equality. The values for the Hubble parameter and matter density parameter will be those obtained by PLANCK \cite{Planck:2018vyg}.

For the condensate $\delta_c$ we will use
\begin{equation}
P_{FDM} (k) = (T_{FDM})^2 P (k) = T_{FDM}^2 (T (k))^2 A_s k^{n_s}
\end{equation}
where the transfer function $T_{FDM}$ is defined as
\begin{equation}
T_{FDM} = \frac{\cos (x^3)}{1 + x^8}
\end{equation}
with $x = 1.61 \left(\frac{m}{10^{- 22} eV} \right)^{1 / 18} \frac{k}{k_{J \ast}}$, and
$k_{J \ast} = 9 \sqrt{\frac{m}{10^{- 22} eV}}$ 1/Mpc, according to \cite{Hu:2000ke}, and used in \cite{Zhang:2018ghp} and \cite{Kulkarni:2020pnb}.

To extract the behaviour of our hybrid model we numerically solve equations \eqref{eqtosol1} and \eqref{eqtosol2} and construct the total overdensity
\begin{equation}
\delta=\frac{\bar{\rho}_c\delta_c+\bar{\rho}_{nc}\delta_{nc}}{\bar{\rho}_c+\bar{\rho}_{nc}}=f\delta_c+(1-f)\delta_{nc}
\end{equation}
for which we compute the power spectrum $P(k)$ and dimensionless power spectrum $\Delta^2(k)$, defined as
\begin{equation}
P(k)=|\delta(k)|^2, \quad \Delta^2(k)=\frac{k^3}{2\pi^2}P(k) \;.
\end{equation}

Based on these, we examine the effect of a variable condensate-fraction $f$, the role of interactions through appropriate choices of $g$ values and how such findings change when we additionally consider the pressure term, parametrised by $\kappa$ of Eq.~(\ref{eq:rhopart}) with $n=5/3$.

\begin{figure}[t!]
    \centering
    \includegraphics[width=0.5\linewidth,]{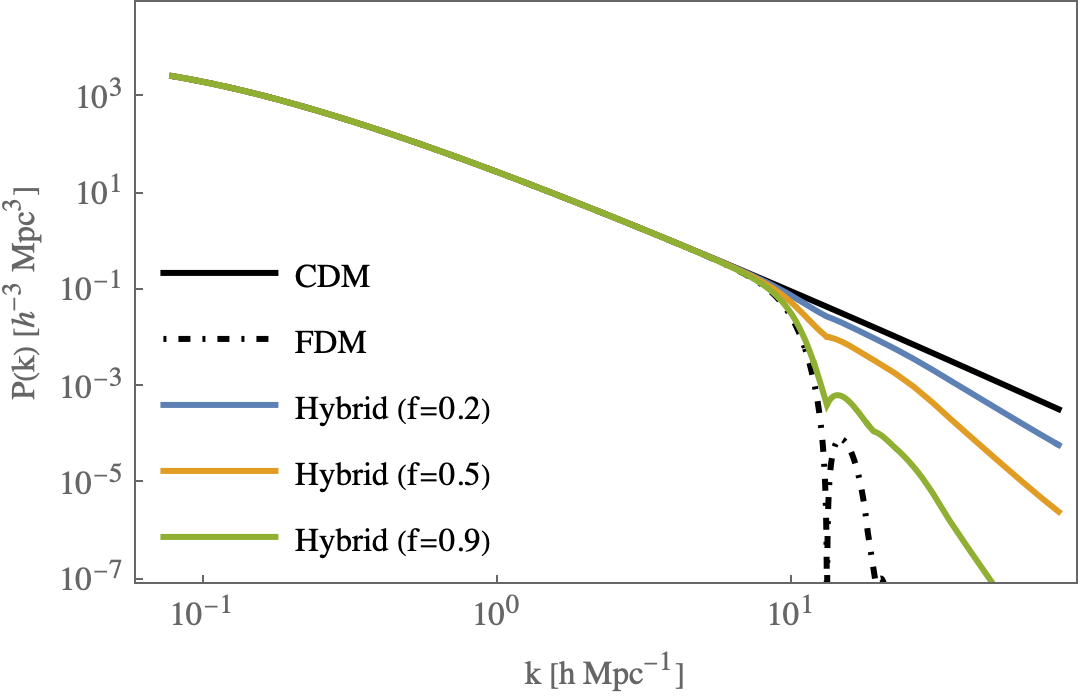}
    \caption{
    Power spectrum in our hybrid condensate-particle dark matter model for different condensate fractions, plotted as a function of $k$ in logarithmic scale for a redshift $z=3$ and a reference boson mass $m=2\cdot 10^{-22}\ eV/c^2$ in the absence of self-interactions ($g=0$) and pressure ($\kappa=0$).
    The established limiting cases of CDM (solid black line, top) and FDM (dot-dashed black line, bottom) are shown for comparison.
    It is clear that increasing the condensate fraction gradually shifts the spectrum downwards, as explicitly shown for the hybrid cases $f=0.2$ (blue), $f=0.5$ (orange) and $f=0.9$ (green). 
    }
    \label{fig:pow1}
\end{figure}

\subsection{Role of Condensate Fraction in Simplest Non-Interacting ($g=0$), Pressure-Less Particle ($\kappa=0$) limit.} \label{sec:4A}

First, we analyse the case of $g=0$ and $\kappa=0$. 
Linearised density perturbations in this context are well known for the cases of CDM, for which the condensate fraction $f=0$ by construction, and have also been explored in FDM (e.g. \cite{Hu:2000ke,Marsh:2010wq,Hlozek:2014lca} which refer to the $g=0$ case), for which $f=1$: these arise naturally as limiting cases of our generalised hybrid equations and are included here as a means of tabulating our model, and for comparison.
Indeed, as shown in Fig.~\ref{fig:pow1}, the hybrid linearised model including both such types of particles through a value of $f \neq \left\{0, 1\right\}$, does indeed smoothly interpolate -- as expected -- between CDM (solid black line) and FDM (dot-dashed black line) values with increasing condensate fraction. This demonstrates the power of our hybrid approach in probing the parameter space between these two established models in the literature.
In this figure, we have used a specific reference value of $m=2\cdot 10^{-22} \ eV/c^2$ for the boson mass, but the conclusions are valid for any boson mass with higher (lower) masses moving the FDM side of the envelope to the right (left). For the aforementioned mass value in the $f\rightarrow 1$ limit there is a sharp reduction of power at about $10 \, h \, Mpc^{-1}$, indicating a characteristic wavelength of about $\lambda =  2\pi/k \simeq \mathcal{O}(1)\, Mpc$ (for a reduced Hubble parameter $h=0.68$) over which quantum pressure is effective in suppressing density fluctuations. This scale reflects a characteristic de Broglie wavelength of the condensed bosons which corresponds to a typical velocity of about $20 \, m\,s^{-1}$. This value will be useful below in estimating the consistency of the $f < 1$ cases.\\

\subsection{Combined Effect of Repulsive Interactions ($g > 0$) and Variable Condensate Fraction $f$ for Pressure-Less Particle Fluid ($\kappa=0$).}

As a natural next step, we consider the effect of self-interactions in our system,  an established generalization of pure FDM \cite{Chavanis_2011a,Chavanis_2011b,Chavanis:2011uv,Fan:2016rda, Suarez:2016eez,Delgado:2022vnt}.
We focus here on the case of repulsive self-interactions ($g>0$) and, to better understand these, we limit the discussion in this section to $\kappa=0$.

\subsubsection{Study of Limiting Cases: $f\to0$ (all particles) and $f\to1$ (all condensate)}

\begin{figure}[t!]
    \centering    \includegraphics[width=0.97\linewidth,]{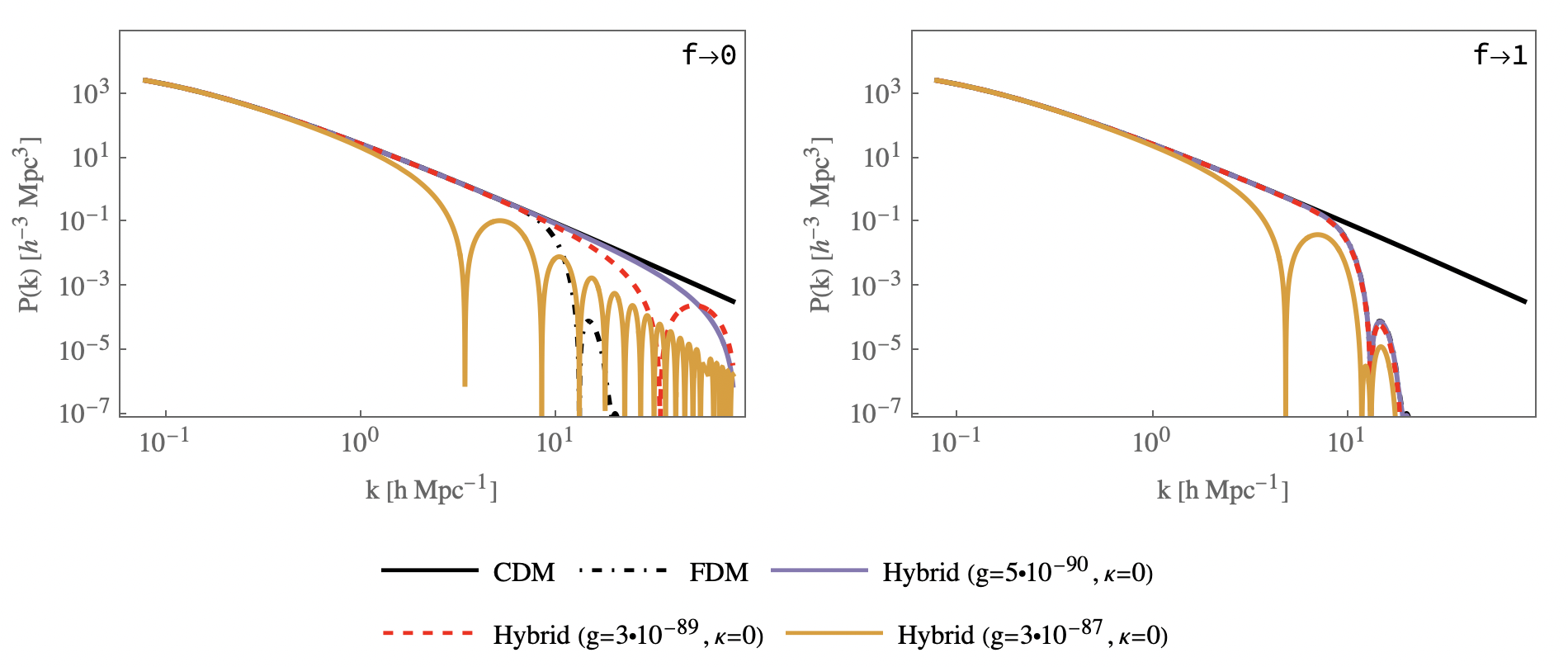}
    \caption{
    Role of interaction strength $g$ on the power spectrum plotted as a function of $k$ in logarithmic scale for a redshift $z=3$ and a reference boson mass $m=2\cdot 10^{-22}\ eV/c^2$ in the absence of pressure ($\kappa=0$) for the cases of purely particle ($f\to0$, left plot) and purely condensate ($f\to1$, right plot) .
    The established limiting cases of CDM (solid black line, top) -- relevant in the $f\to0$ case -- and FDM (dot-dashed black line, bottom) -- correspondingly relevant in the $f\to1$ limit -- are shown for comparison in both plots. In each case we see a shift of the power spectrum to lower values of $k$ without any identified lower bounds, and the emergence of an oscillatory behaviour, with the spectrum bounded from above by the CDM ($f\to0$ case) and FDM ($f\to1$ case) respectively.
    Left: For $f\to0$ the power spectrum of the interacting case is oscillatory at large $k$ with a relatively slow decrease in amplitude. 
    Right: for $f\to1$, the limiting behaviour at the interacting model at high wavenumbers is that of FDM with a relatively sharp cutoff and no short scale oscillations. 
    }
    \label{fig:pow2}
\end{figure}

We first consider the limit $f\to0$, which is the situation where all the system is non-condensed and, for reference, we use the same mass $m=2\cdot 10^{-22} \ eV/c^2$. 
As we increase the interaction strength $g$ to positive values, we find that
the power spectrum starts deviating, and dropping below (i.e.~bounded upwards by) the CDM one (solid black line) at large $k$ values, while also exhibiting oscillations: such behaviour is shown for various interaction strengths $g$ in the left plot of Fig.~\ref{fig:pow2}. The deviation point from the CDM curve moves to lower $k$ and the spectrum exhibits more frequent oscillations with increasing $g$, as expected from a perusal of Eq.~\eqref{eqtosol2} in this limit. 
An oscillatory decaying behaviour of the power spectrum, but with a much sharper cutoff due to the quantum pressure contribution,  has been previously noted in the case of (non-interacting) FDM \cite{Hlozek:2014lca,Marsh:2015xka} (which would amount here to the $f\to1$ limit): this is shown here for comparison by the dot-dashed line. Interestingly, although for small $g>0$ values the power spectrum lies above that of the FDM (for which $g=0$), further increasing the repulsive interaction strength can lead to the deviation from CDM occurring at wavenumbers even smaller than those set by the FDM's ($g=0$) quantum pressure -- an example of this for a value of $g=3\cdot 10^{-87} J m^3$ is shown by the orange curve in the left plot in figure \ref{fig:pow2}.

\begin{figure}[t!]
    \centering  \includegraphics[width=0.97\linewidth,]{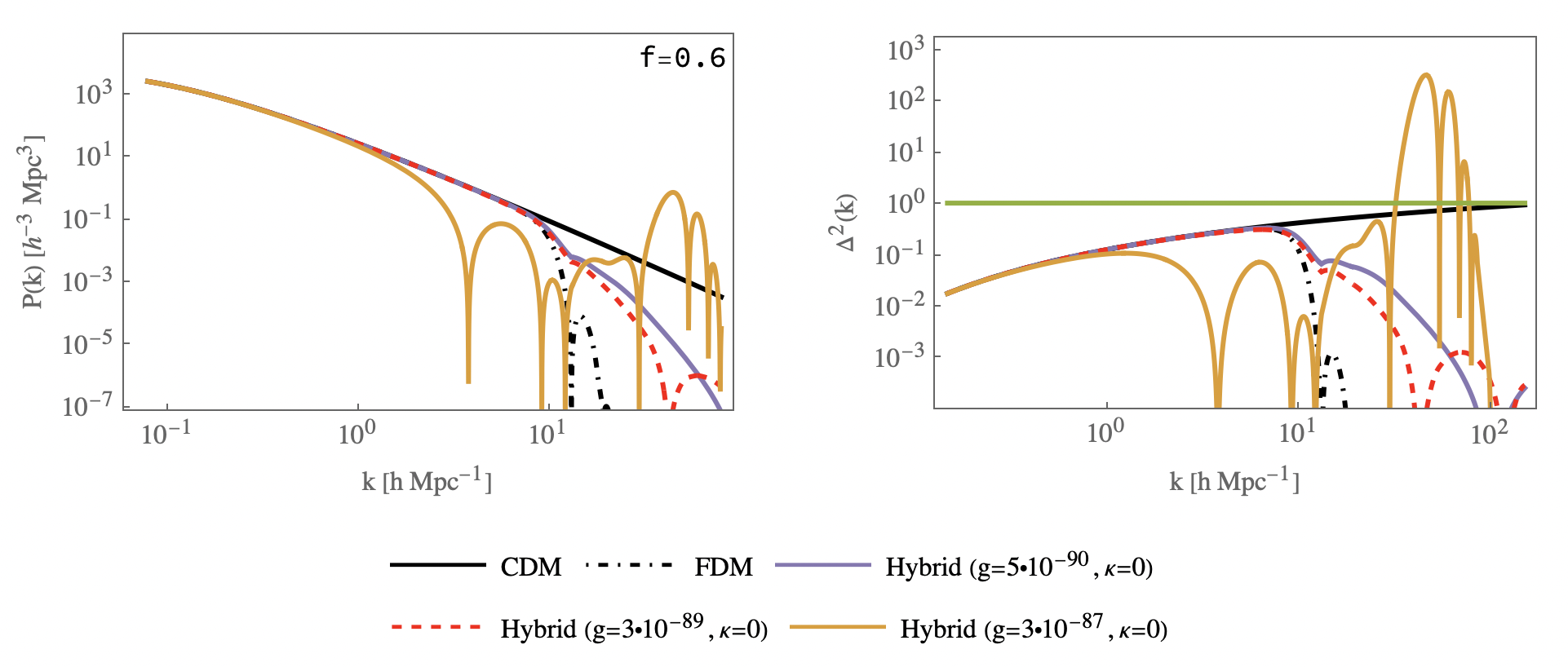}
    \caption{
    Left: Typical power spectrum example, similar to Fig.~\ref{fig:pow2}, but now in the simultaneous presence of both condensed and non-condensed components, shown here for the specific case of $f=0.6$ (and, still, $\kappa=0$), $m=2\cdot 10^{-22} eV/c^2$ and $z=3$. While the generic features discussed previously at smaller $k$ still persist, we see here that for sufficiently high $g$ (orange curve), the hybrid power spectrum can even exceed, at high $k$, the CDM spectrum.
    Right: Corresponding dimensionless power spectrum as function of $k$ in logarithmic scale. The green line defines the limit between the linear and non-linear regimes. Interestingly for sufficiently high $g$ (orange curve), the spectrum even enters into the non-linear regime where the presented perturbation theory breaks down, indicating that the co-existence of condensed and non-condensed particles could potentially enhance smaller-scale structure (although our present analysis cannot describe such regime which would require fully non-linear numerical approaches).}
    \label{fig:boundmg}
\end{figure}

Moving now to the opposite, $f\to1$ limit, which corresponds to having all bosons in the condensate, we see that it is now the FDM case that provides an upper bound for the power spectrum of the interacting system.
In other words, in both limiting cases considered in the appropriate limits ($f\to0$, left image vs.~$f\to1$, right image), it is actually the relevant model (CDM, FDM) which sets an upper bound to the power spectrum, with no lower bound provided. For example, also in the $f\to1$ case we see that for values of $g>3\cdot 10^{-89} J m^3$, the power spectrum again deviates from the CDM line at a wavenumber to the left of the FDM scale (orange curve in the right plot in figure \ref{fig:pow2}). 

To summarize, these findings imply that (at least in the $\kappa = 0$ limit) the power spectrum predictions in either CDM or FDM provide an \emph{upper bound} to the $g\neq 0$ power spectrum, with increasing self interaction driving the power below the CDM/FDM power, adding further oscillations with increasing $k$ and moving the point of deviation from the CDM/FDM spectra to lower values of $k$.

\subsubsection{Case $0<f<1$}

\begin{figure}[b!]
    \centering  \includegraphics[width=0.98\linewidth,]{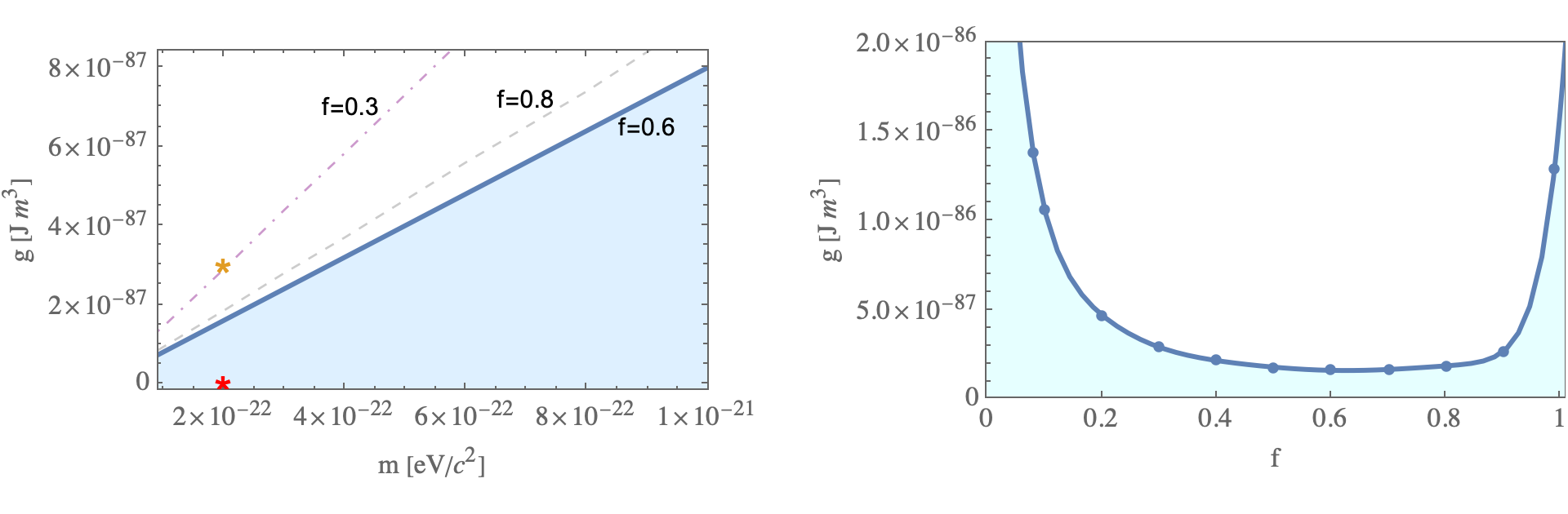}
    \caption{
Left: Characterisation of the parameter range in $g-m$ space for which the power spectra of hybrid models lie, for a given condensate fraction $f$, consistently below the CDM curve for all (probed) $k$ values: the blue shaded region identifies such regime of CDM-bounded spectra for the $f=0.6$ case of Fig.~\ref{fig:boundmg}, with the solid blue line showcasing the locus of all those $(m,\,g)$ values for which the power spectrum touches the CDM curve for at least one value of $k$ (at $z=3$). The $(m,\,g)$ values corresponding to the red and orange curves shown in Fig.~\ref{fig:boundmg} are indicated by the like-coloured stars in this plot, which are respectively located below and above the blue line. 
The value of $f=0.6$ has been chosen as a conservative estimate, with higher and lower condensate fractions leading to even broader parameter range, identified as the regions below the corresponding lines indicated by $f=0.8$ (grey dashed) and $f=0.3$ (purple dot-dashed).
Right: Corresponding $g-f$ parameter region for which the hybrid model power spectrum always lies below the CDM one, shown here by the cyan-shaded region for a fixed mass $m=2\cdot 10^{-22} eV/c^2$; the existence of a local minimum around a condensate fraction of $f\simeq 0.6$, explains such choice of $f$ on the left (and other related) plots, as for such values the enhancement in power occurs for the smallest values of $g$. In particular, we note that the curve increases monotonically in the limits $f\to 0$ and $f\to1$, indicating that all power spectra lie below the CDM ones in such limits, as expected. The blue dots indicate the values of $f$ that we have examined and the blue line separating the two regions of the parameter space is constructed via interpolation.}
    \label{fig:boundmg2}
\end{figure}

We now turn our attention to condensate fractions between the two extreme cases. Firstly, we note that the shift of the deviation of the power spectrum from the CDM one to smaller $k$ values with increasing $g$, and its oscillatory features, continue in this regime, again being unbounded from below even in the presence of coupling between the two components (condensate, particles).
Our findings indicate that, for a given $g$ (and $m$), the value of $k$ at which the spectrum starts bending away from the corresponding CDM one, is not significantly modified by a variation in the condensate fraction $f$. Instead, the main effect of $f$ is to modify the initial inclination of the curve after it deviates from the CDM one, pushing it towards the CDM case for $f\to0$ or towards the FDM curve for $f\to1$. 
%
More interesting, however, is the observation that,
for a fixed mass, and for big enough values of $g$, a condensate fraction between $0$ and $1$ leads to an enhancement in the power spectrum for a range of small scales such that it grows beyond the CDM case, entering quickly into the non-linear regime. 
To see this effect we refer the reader to the orange curve in figure \ref{fig:boundmg}, where the curves correspond to a redshift $z=3$. For illustration we demonstrate the case $f=0.6$ which we also found to be the approximate value which makes the effect maximal (see subsequent figure \ref{fig:boundmg2}). Within this linearized model, there is an enhancement of power which exceeds the CDM result in a band of wavenumbers within the oscillatory regime. To highlight this more clearly, in the right plot of figure \ref{fig:boundmg} we show the dimensionless power spectrum $\Delta^2(k)$ where the green line defines the limit of the linear regime: for linear perturbation theory to be trustworthy the power spectrum must lie below the green line for which $\Delta^2=1$. Clearly, perturbation theory cannot be valid here and a fully non-linear treatment is required. However, the fact that the power spectrum is enhanced into the non-linear regime for situations where standard CDM is not would imply that non-linear structures could be formed early in the universe at such scales for such hybrid models.

This enhancement is negligible if the condensate fractions are closer to almost all-condensate or all-particles, showing that it is generated by the dynamical interplay of the two coupled components.
%
The shaded area in the left plot of figure \ref{fig:boundmg2} designates the region of values for $g$ where the hybrid curve is always bounded by the CDM curve for a range of boson masses between $10^{-22} \ eV/c^2$ to $10^{-21} \ eV/c^2$, considering $f=0.6$, close to which we have found the enhancement to be strongest.
For comparison, the purple and gray dashed lines indicate where the limit of such a region would be for $f=0.3$ and $f=0.8$ respectively, showing that any possible bounds on $g$ are always above the ones found for $f=0.6$, as we point out in the right plot of figure \ref{fig:boundmg2}.

\subsection{Role of Particle-Component Pressure ($\kappa\neq 0$) with/without Self-Interactions}

The particle component will be comprised of particles with higher velocities than the condensed component which would be expected to have a non-zero kinetic pressure, stemming from velocity dispersion. We have modelled this pressure here with the equation of state \eqref{eq:rhopart} and now turn to the effects of such a non-zero pressure, discussing it in the non-interacting limit and for both repulsive and attractive self-interactions.

\subsubsection{Non-Interacting Limit ($g=0$) with $\kappa \neq 0$} 

\begin{figure}[b!]
    \centering    \includegraphics[width=0.98\linewidth,]{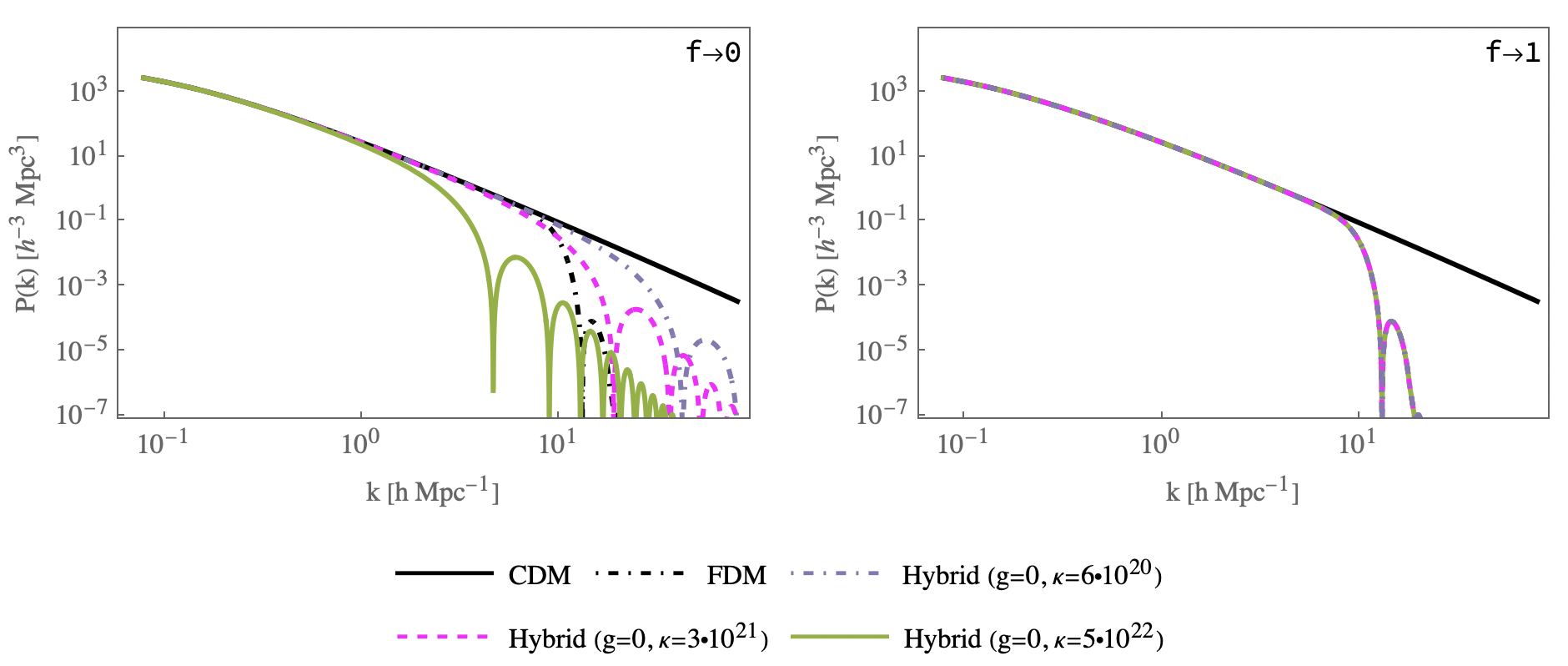}
    \caption{
    Role of particle pressure $\kappa$ on the power spectrum plotted for three different values of $\kappa$ (see figure key) as a function of $k$ in logarithmic scale for a redshift $z=3$ and a reference boson mass $m=2\cdot 10^{-22}\ eV/c^2$ for a non-interacting system ($g=0$) for the cases of purely particle ($f\to0$, left plot) and purely condensate ($f\to1$, right plot).
Left: In the $f\to0$ limit, there is a shift of the power spectrum to lower values of $k$ without any identified lower bounds, and the emergence of an oscillatory behaviour, with the spectrum bounded from above by the CDM case.
Right: In the $f\to1$ limit all $\kappa \neq 0$ spectra overlap exactly with the FDM prediction. 
    }
    \label{fig:pow4}
\end{figure}

To isolate the effect of the particle pressure, we first discuss the power spectra arising in the non-interacting limit ($g=0$). We first highlight the limiting cases $f\to0$ and $f\to1$, which are shown in Fig.~\ref{fig:pow4} for a mass of $m=2\cdot 10^{-22} \ eV/c^2$.
The left plot shows that increasing particle pressure in the $f\to0$ limit has a qualitatively similar effect as was previously found (Fig.~\ref{fig:pow2}) for increasing interaction, i.e.~a decrease of the power spectrum compared to the CDM limit (solid black line) at high $k$, and an oscillatory behaviour, with higher $\kappa$ amplifying the effect and enhancing the presence of oscillations. The similarity between the effects of varying $g$ and varying $\kappa$ in the $f\to0$ limit is expected from a perusal of Eq.~\eqref{eqtosol2}.\footnote{The terms proportional to these two parameters are not identical however: since they carry different factors of $a$, the $g$ term is more dominant than the $\kappa$ term at earlier times. We leave a more detailed study of the redshift dependence of perturbations in our model for future work.} In particular, for larger values of $\kappa$, the spectrum deviates from the corresponding CDM one at a lower $k$ value than the corresponding FDM ($g=0$, $\kappa=0$, dot-dashed line) prediction -- this can be seen by the green curve in Fig.~\ref{fig:pow4} for a value of $\kappa = 3\cdot 10^{21} \ \ {m^4}{s^{-2} kg^{-2/3}}$. Since the $\kappa$ terms become more dominant at later times, they will eventually lead to a steeper decay of the oscillations at a high $k$, something we do observe for our probed values. 

The inclusion of $\kappa$ has been on phenomenological grounds (see section \ref{IIC} for the relevant discussion) and we do not have an {\em a priori} value for this parameter without further knowledge of the underlying phase space distribution. In order to estimate the order of magnitude of $\kappa$ for which the relevant terms can have an effect at the redshifts we consider here, we can directly compare with the values of $g$ used through the text from equation \eqref{eqhyd4b}. Imposing that the particle-component pressure is comparable with that coming from the self-interaction around $z=0$, we must have that
\begin{equation}
\frac{g}{m^2} \tilde{\rho} \sim \kappa \tilde{\rho}^{2 / 3}
\end{equation}
Considering our reference values of $m=2\cdot 10^{-22} eV/c^2$, $g$ of the order of $10^{-87} J m^3$ and as reference density the local dark matter density $0.0122\, {M_{\odot}}/{pc^3}$ \cite{Sivertsson:2017rkp}, the order of magnitude of $\kappa$ is around $10^{21} \ {m^4}{s^{-2} kg^{-2/3}}$. Using \eqref{eq:typ_vel}, this corresponds to a typical velocity of about $5000 \, m s^{-1}$. The typical de Broglie wavelength of such a population of particles is more than two orders of magnitude shorter than that of the component described by $\Phi_0$ ($k\simeq 10\, h\,Mpc^{-1}$, see section \ref{sec:4A}), hence validating the use of a phase space distribution at the length scales we are analysing ($k<10^2 h\,Mpc^{-1}$).

For the case $f\to1$ (all condensate and absence of particles) a non-zero $\kappa$ plays no role since the particle pressure is not present in the condensate equation. This is shown in the right plot of figure \ref{fig:pow4}, giving the same FDM curve for any choice of $\kappa$, as expected.

\subsubsection{Repulsive Interactions ($g > 0$) with $\kappa\neq 0$}

Given the qualitative similarity of an increasing $g>0$ (with $\kappa=0$) [Fig.~\ref{fig:boundmg}] and increasing $\kappa>0$ (with $g=0$) [Fig.~\ref{fig:pow4}] in the $f\to0$ limit, it is natural to see how the two compare and -- more importantly -- to examine what would be the combined effect of $g>0$ and $\kappa>0$ for a general intermediate $0<f<1$ value. We discuss this in the context of our previously-considered $f=0.6$ representative case.

\begin{figure}[h!]
    \centering    \includegraphics[width=0.5\linewidth,]{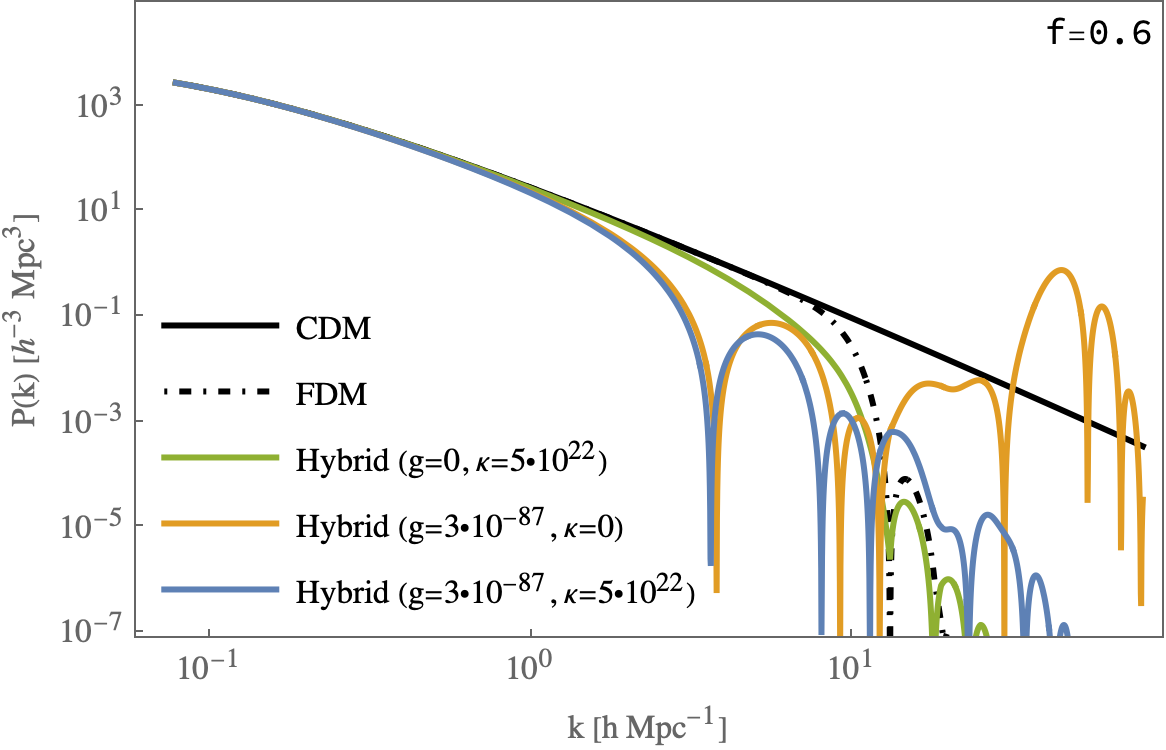}
    \caption{
 Interplay of repulsive interactions ($g>0$) and particle pressure $\kappa>0$ on the power spectrum for $f=0.6$ for a redshift $z=3$ and a reference boson mass $m=2\cdot 10^{-22}\ eV/c^2$.
 The addition of a non-zero $\kappa$ term can alleviate the large $k$ interaction-induced increase of the hybrid power spectrum beyond the corresponding CDM one (orange plot), as evident from the blue curve.
 }
    \label{fig:pow5}
\end{figure}

Interestingly, the previously-discussed enhancement of the power spectrum at large $k$ for large $g>0$ and $f=0.6$ [Fig.~\ref{fig:boundmg}, orange curve] seems to be largely alleviated by the addition of the particle pressure ($\kappa \neq 0$): this is evident in Fig.~\ref{fig:pow5}, comparing the hybrid power spectra for $f=0.6$ with $g=0$, $\kappa \neq 0$ (green), $g \neq 0$, $\kappa =0$ (orange) and $g \neq 0$, $\kappa \neq 0$ (blue). We observe that the existence of pressure in the particle component (at least modelled as in Eq. \eqref{eq:rhopart}) reduces the previous excess and the spectrum is now again bounded by the CDM curve. Comparing the blue curve to the green curve, representing a case with pressure but no self-coupling for the particles, we see that the introduction of a non-zero $g>0$ is to enhance the perturbations on small scales, somewhat countering the  effects of the particle pressure.  \\

\subsubsection{Attractive Interactions ($g<0$) with $\kappa\neq0$}

So far we have considered a positive value of $g$. However, an attractive self-interaction ($g<0$) can appear in axion contexts and already studied for example in \cite{Chavanis_2011a,Chavanis_2011b,Chavanis:2016dab,Desjacques:2017fmf,Mocz:2023adf}. In particular, it was shown in \cite{Desjacques:2017fmf} that even a tiny attractive self-interaction could be important in the stability of structures, and numerical studies have been carried out recently \cite{Mocz:2023adf}.

\begin{figure}[t!]
    \centering    \includegraphics[width=0.5\linewidth,]{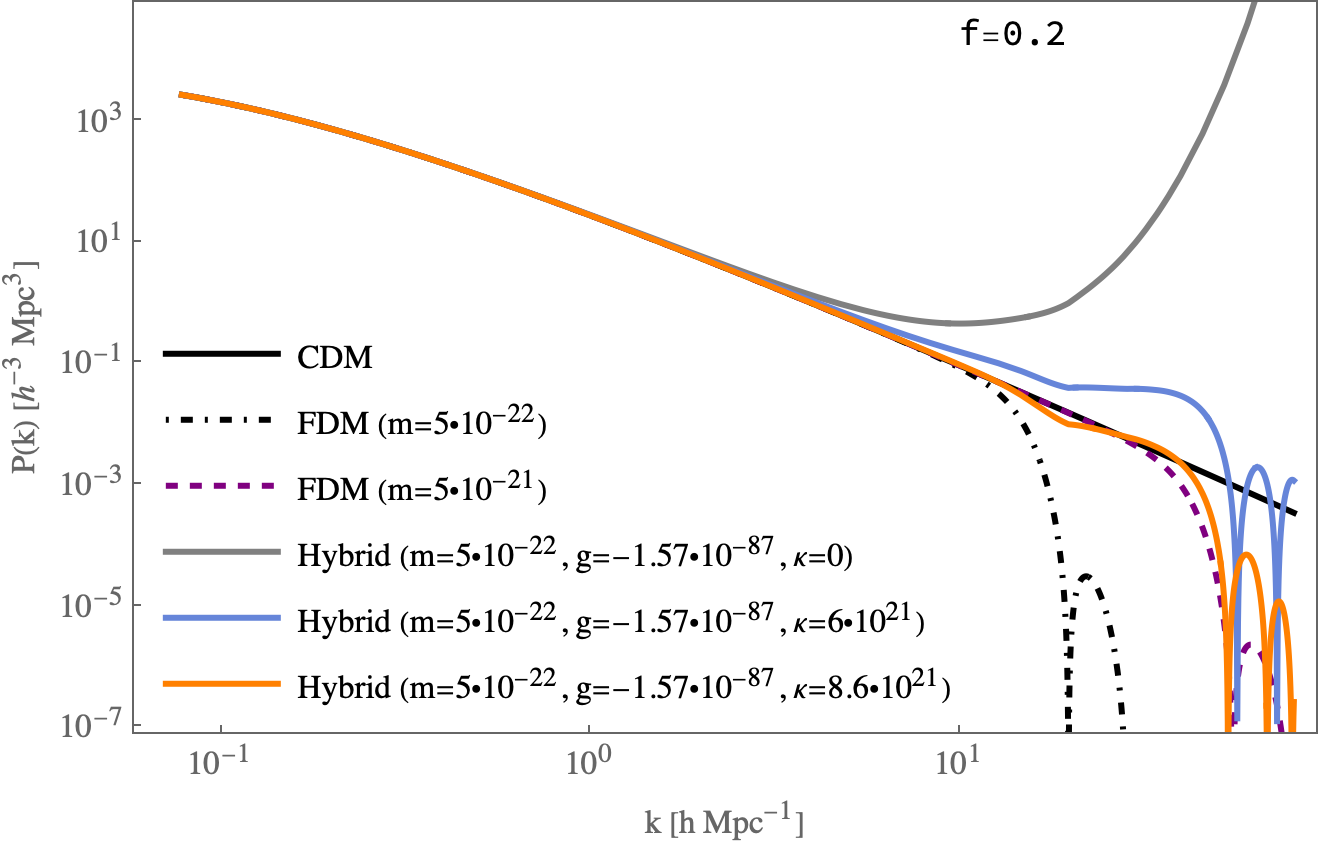}
    \caption{
Interplay of particle mass, attractive interactions ($g<0$) and particle pressure $\kappa>0$ on the power spectrum for a small condensate fraction of $f=0.2$ at a redshift $z=3$. The presence of attractive interactions without particle pressure (grey line) sends the system into the nonlinear regime for sufficiently large $|g|$, significantly exceeding the CDM line (black). The addition of increasing values of particle pressure $\kappa$ counterbalances this effect (blue and orange lines with increasing $|g|$ and $\kappa \neq 0$). Remarkably, the combined effect of modest attractive interaction ($g=-1.57\cdot 10^{-87} J m^3$) and particle pressure ($\kappa=8.6\cdot 10^{21} \ {m^4}{s^{-2} kg^{-2/3}}$) for a particle of mass $m=5\cdot 10^{-22} \ eV/c^2$ (orange curve) leads to effectively the same power spectrum as that of FDM (for which $g=\kappa=0$) but for a particle mass one order of magnitude higher ($m=5\cdot 10^{-21} \ eV/c^2$, purple dashed line).
}
    \label{fig:pow6}
\end{figure}

As an example, we consider a value of $g=-1.57\cdot 10^{-87} J m^3$, corresponding to one of the values for the scattering length used in \cite{Mocz:2023adf} and a mass of $m=5\cdot 10^{-22} \ eV/c^2$, along with a condensate fraction of $f=0.2$. 
As pointed out in \cite{Desjacques:2017fmf}, negative values of $g$ for a pure condensate case ($f=1$) can lead to instabilities and we observe this in our analysis even in the hybrid case:  see the rapid growth of power at small scales in fig.~\ref{fig:pow6} (grey line) which implies a quick transition to the non-linear regime at high $k$. This would of course call for a full non-linear treatment before any valid conclusions could be reached as the presented linearized analysis is obviously no longer valid.
With the addition of the particle pressure however, values of $\kappa$ can be chosen so that the destabilising effect of a negative $g$ can be counteracted, keeping the spectrum in the linear regime, as indicated by the orange curve in fig \ref{fig:pow6}. 

Interestingly, adjusting the particle pressure can result in a curve that is very close to a typical FDM curve but for a \emph{heavier} mass of $m=5\cdot 10^{-21} \ eV/c^2$. With this in mind, the existence of these two competing effects can lead to a \emph{parameter degeneracy} that could bypass mass constraints as those presented in \cite{Irsic:2017yje}, where the boson mass is estimated as $m>10^{-21} \ eV/c^2$ with Ly-$\alpha$ forest data, but without the consideration of self-interactions and the existence of a particle component, or the higher bounds using Ly-$\alpha$ obtained in \cite{Rogers:2020ltq}, where $m>2\cdot 10^{-20} \ eV/c^2$. In any case, our analysis here shows that the particle pressure would allow to have bosonic masses below such bounds but with similar behaviour to FDM models of higher masses. Finally, it is interesting to note that the balance between a negative self-coupling and the pressure in the particle sector can also lead to localised increases of power that could push some scales into the non-linear regime earlier - see the blue line in fig.~\ref{fig:pow6} .\\

\section{Conclusions}\label{sec5}

In this work we have examined the behaviour of linear density perturbations in a model of dark matter that consists of both an effectively condensed, coherent part, composed of low velocity modes and governed by the Gross-Pitaevskii equation, as well as a non-condensed, incoherent part, composed of faster moving particles and described by a Boltzmann equation. The system is self-gravitating in an expanding universe with Hubble and density parameters set at the Planck satellite values, and is also self-interacting with a quartic interaction with a self-coupling constant $g$. The full equations for such a hybrid model were derived from first principles from the non-relativistic,  bosonic action in \cite{Proukakis:2023nmm}, and our present work analysed the hydrodynamical limit, including mean-field effects in a self-consistent manner to order $g$, while ignoring collision terms of order $g^2$.
Moreover, in order to close the Boltzmann moment equations, we
 made some simplifying assumptions. We further assumed that the pressure due to velocity dispersion in the particle component can be modeled via \eqref{eq:rhopart} which allowed us to further drop one of the resulting hydrodynamical equations. 

This hybrid model reproduces CDM and fuzzy dark matter (FDM, with $g=0$) in the respective limits $f\to0$ and $f\to1$ of the condensate fraction $f$. Inclusion of a repulsive self-interaction ($g>0$) introduces damped oscillations on small scales in the particle-only limit. Similar oscillations exist for $f\to1$ for sufficiently large $g$ but the behaviour tends to that of FDM, with a sharp cutoff for small $g$. An interesting behaviour arises for mixtures where both components are present and comparable: a repulsive self-coupling can couple the perturbations in the two components leading to localized enhancement in the power spectrum that can even push perturbations into the non-linear regime at  times earlier than those expected from $\Lambda CDM$. However, such enhancements can be moderated by the inclusion of pressure in the non-condensed,  particle component, at least if this pressure is modeled by \eqref{eq:rhopart}. 
The strength of such effects naturally depends on the choice of model parameters and a better understanding of any such closure conditions for the Boltzmann hierarchy is required before any definite conclusion can be drawn on the existence, importance and impact of $\kappa \neq 0$ terms.    

We also considered an attractive self-coupling ($g<0$) and observed that it induces instability and perturbation growth at small scales. Such instabilities for pure condensate states were already pointed out in \cite{Desjacques:2017fmf} and we observe similar behaviour even at a relatively low condensate fraction of $f=0.2$. The addition of pressure on the incoherent particles, modelled here by $\kappa \neq 0$, can counteract such instability. An interesting consequence is that the combined effect of the self interaction and the pressure can result in a spectrum that resembles that of pure FDM ($f=1$, $g=0$, $\kappa=0$) but of a \emph{higher boson mass} which, for the values of interaction and pressure presumed here, could amount to an increase of approximately one order of magnitude. Hence, mixtures of condensed and particle components, like the ones described here, introduce degeneracies that should be kept in mind when placing observational limits on the boson mass. Furthermore, particle pressure can result in localized features of enhanced power at specific wavenumbers, depending on the choice of parameters. The result of such enhancements can only be assessed via fully non-linear hybrid simulations based on our full model of Eqs.~(\ref{initial1})-(\ref{initial3}). 

We note here that the extent of hybridisation between purely incoherent particles and purely coherent waves was parametrised by an effective (constant) condensate fraction parameter.
In a homogeneous system in thermodynamic equilibrium,
such a quantity is directly constrained by thermodynamic considerations dictating the particle distribution function, and precise knowledge of
relevant parameters (e.g. interaction strength, total density).
However, the form of such a distribution  in our {\em quasi-static} gravitating scenario is not {\em a priori} known, and the local condensate fraction is also expected to vary both spatially and temporally: such factors make it difficult to constrain the mean value of the condensate fraction in our system, and this is the reason why this work characterised the system properties treating the effective condensate fraction as an unconstrained variable.
These are nonetheless important questions which need to be addressed by direct numerical simulations of Eqs.~(\ref{initial1})-(\ref{initial3}), which is left to future work.  

Bosons can behave both as incoherent collections of particles and also as coherent waves. This work examined hybrid states where both behaviours can coexist and showed that bosonic dark matter could have a rather rich phenomenology in terms of the matter power spectrum, not necessarily limited to the wavy effects of what is usually referred to as fuzzy dark matter.  This phenomenology depends in turn on whether such states can be realized in the universe and, ultimately, the dark matter production mechanism for which we can only be agnostic at the moment.
It is worth noting here that our model bears some similarities to models of mixed dark matter that have started being investigated in the literature \cite{Schwabe:2020eac, Vogt:2022bwy} but with the important difference that these works consider different particle species, with correspondingly different masses. In our case, self-interactions play an important role and they are naturally inherited by both components which are now coupled directly, beyond the mere effect of gravity. It is such coupling that gives rise to the phenomenology we discuss here. 
Formally however, the equations are very similar and one might further envisage couplings even between different species. Such models would have comparable phenomenology, depending on the form and values of these couplings and hence the analysis and conclusions of this work might be useful even in the context of more general, mixed dark matter models with couplings between the different species. Examples might involve species with different masses, one of which is described by a wavy condensate and another by a phase space distribution of particles. Finally, when the phenomenology we examined here is taken into account, it would affect any parameter estimation from observations, not least due to the existence of degeneracies as we have alluded to above. Further investigation is therefore required, especially in understanding the corpuscular, collisional component, before observations can safely place limits on generic models of bosonic dark matter.

\section*{Acknowledgements}
This work is supported by the Leverhulme Trust, Grant no.
RPG-2021-010. 



\bibliographystyle{unsrt}
\bibliography{Refs_MixedDM}

\end{document}